\documentstyle[graphicx]{WORKSHOP}
\begin{document}

% TITLE OF THE PAPER
%  If the title is too long for a single line, you can split it 
%  by putting two backslashes. 
%  You might want to put the subtitle. Then it should be inserted 
%  within {\large\sf  }.
%  e.g.:  
%     \title{ Too Long Title \\ for one line \\
%     {\large\sf Subtitle} }
\title{
Scaling accretion flow models from BHB to AGN \\ 
{\large\sf  -- Why doesn't it work? --} % SUBTITLE
}

% AUTHOR(S) 
\author{
Chris Done$^1$
\\[12pt]  % TO BE SPACED WITH ONE LINE
%
% INSTITUTES OF AUTHORS
$^1$  Department of Physics, University of Durham, South Road, Durham
DH1 4ED  \\
%
% please put the first author's initial and e-mail address below
{\it E-mail(CD): chris.done@durham.ac.uk} 
%            \_ Initial      \
%                             \_ E-mail address
}

\abst{ 
Black holes depend only on mass and spin, while what we see
from the accretion flow in steady state depends also on mass accretion
rate and (weakly) inclination. Hence we should be able to scale the
accretion flow properties from the stellar to the supermassive black
holes. But the data show significant differences between these two
types of systems, suggesting that we are missing some crucial physics
in AGN.  One of these differences is the soft X-ray excess which is
seen ubiquitously in bright AGN, but only occasionally in BHB.
Another is the much faster variability seen in the high energy tail of
high mass accretion rate AGN compared to that seen in the tail of
BHB. We show that while this variability is not understood, it can be
used via the new spectral-timing techniques to constrain the nature of
the soft X-ray excess. The coherence, lag-frequency and lag-energy
results strongly support this being an additional low temperature
Comptonisation component rather than extreme relativistically smeared
reflection in the 'simple' Narrow Line Seyfert 1 PG1244+026.  
}

\kword{Accretion flows: Black holes: Binaries: Active Galactic Nuclei}

\maketitle
\thispagestyle{empty}

\section{Introduction}

Understanding accretion in strong gravity is a challenge, so we need
to use all available information to build up a physical picture of
what happens. The black hole binaries (BHB) give an observational
template of how the accretion flow varies as a function of (mostly)
$\dot{m}=L/L_{Edd}$, which we can then scale up to the supermassive
black holes in Active Galactic Nuclei (AGN) to predict how we expect
AGN to behave. Any differences between these models and the data mean
that there are some aspects which do not simply scale with mass,
giving insight into the additional physical processes required.

\section{Black hole Binaries}

In BHB the overall spectral and timing properties can all be fit into
a model where the standard disc extends down to the last stable orbit
only at high $L/L_{Edd}$. At luminosities below $L\sim 0.02L_{Edd}$
the optically thick, geometrically thin inner disc progressively
receeds, probably via evaporation (Meyer \& Meyer-Hofmeister 1994;
Mayer \& Pringle 2007) and is replaced by a hot, probably radiatively
inefficient, optically thin, geometrically thick flow.  The
characteristic hard X-ray spectrum and variablity are (mostly)
generated in the flow, with the longest variability timescales set by
the outer radius of the flow. As the disc receeds, this outer flow
size increases, so all the timescales increase (including the QPO
which may be a sign of vertical precession of the entire hot flow:
Ingram \& Done 2012). The drop in disc seed photons means that this
correlates with a decrease in direct disc emission, and with
increasing spectral hardness of the high energy tail (see e.g. the
reviews by Remillard \& McClintock 2006; Done, Gierlinski \& Kubota
2007).

Transition spectra are complex, with strong disc and strong coronal
emission. The corona is also probably inhomogeneous, with softer
compton spectra emitted in the outer regions of the flow which is
closer to the disc, and harder spectra in the inner regions which
intercept less disc emission. Signatures of this are that the compton
emission has complex time lags associated with it. Photons at higher
energies are correlated with those at lower energies, but with a
(frequency dependent) lag time (Miyamoto \& Kitamoto 1989; Nowak et al
1999). This can be modelled by each part of the flow generating
variability on a timescale which decreases with radius, and where
variability at each radius propagates down through the accretion flow
to modulate the variability at smaller radii (Kotov et al 2001;
Arevalo \& Uttley 2006; Ingram \& Done 2012; Axelsson et al 2014)

At very high luminosities the spectra can also be similarly complex,
with a strong disc and strong soft compton tail (Belloni et al 2005).
These 'very high' or 'steep power law' state spectra may also
represent ones where the disc is slightly truncated (Done \& Kubota
2006; Tamura et al 2012).  However, evaporation is no longer
sufficient to remove the inner disc at these high mass accretion
rates, so radiation and/or magnetic pressure (Meier 2005; Machida et
al 2006) may play a role.

\section{Scale to AGN}

AGN live in more complex environments than BHB, so we select only
those where obscuration is not an issue. However, scaling from BHB
predicts that the intrinsic spectral energy distribution should change
similarly dramatically with $L/L_{Edd}$, unlike the simplest unified
AGN models in which obscuration is the only determinant of the
observed spectrum. Another difference between AGN and BHB is that the
BHB only span a small range in black hole mass (less than a factor of
2) so they form a very homogeneous sample, whereas AGN range from
$10^5-10^{10}M_\odot$. Hence we expect the unobscured AGN spectra to
change with both mass and $\dot{m}=L/L_{Edd}$, most obviously as the
accretion disc should peak at $T_{disc}\propto
(\dot{m}/M)^{1/4}$. This means that the disc typically peaks in the
(often unobservable) UV-EUV, so we need multiwavelength data rather
than simply using X-ray data to constrain the disc and tail
simulateously as in BHB.

Starting from the BHB, we might then match up some of the different
unobscured AGN types to the BHB states. LINERs are at low $L/L_{Edd}$
so may be the analogue of the low/hard state with the inner disc
replaced by a hot flow. Seyferts span the transition between the
brightest low/hard states and disc dominated states.  Quasars are at
similar $L/L_{Edd}$ as Seyferts, but with a higher mass black hole
giving their higher luminosity. And perhaps the Narrow Line Seyfert
1's are the analogue of the very high state in BHB since these are
known to be at high $L/L_{Edd}$ (e.g. Boroson 2002; Woo \& Urry
2002). Some evidence for this is that the bolometric correction from the
X-ray flux increases substantially with $L/L_{Edd}$ in AGN in a
similar way to BHB (e.g. Vasudevan \& Fabian 2007).

\begin{figure}[!t]
\centering
%\psbox[xsize=0.4#1,ysize=0.2#1,rotate=r]
\includegraphics[width=8cm]{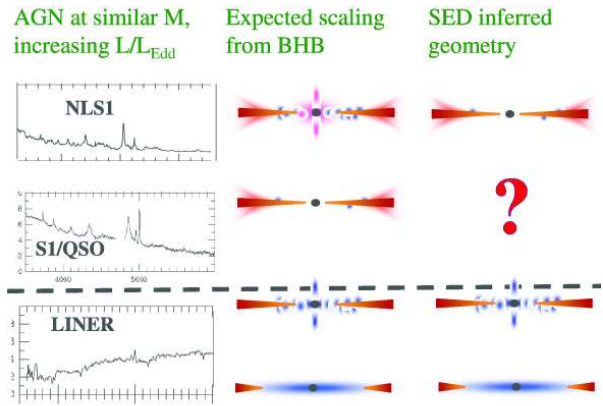}
\caption{(left panel) AGN at similar mass but different $L/L_{Edd}$
  have different optical line ratios, signalling a change in their
  SED. (middle panel) The geometries inferred for BHB as a function of
  increasing $L/L_{Edd}$, with the dashed black line indicating the
  collapse of the hot inner flow. (right panel) The observed AGN SED show that
  while LINERs may well have the truncated disc/hot inner flow of
  low/hard state BHB, standard broad line Seyferts and Quasars do
  not look like the disc dominated (or in any other) BHB state. Narrow
  Line Seyfert 1s at high $L/L_{Edd}$ might be expected to be
  analogous to the BHB very high state, but their spectra look more
  like the disc dominated state!  }
\end{figure}

However, compiling the SED's of these different $L/L_{Edd}$ AGN does
not generally support this picture. The only one which seems to work
is the LINERs as the analogue of the low/hard state (though even this
is somewhat controversial e.g. Maoz 2007 vs Nemmen et al 2014).  The
lack of an inner disc means no strong UV emission so no broad line
region, explaining the optical line ratios which define this class.

But classic Quasar spectra are not disc dominated (as has been known
for years, e.g. Elvis et al 1994). Rather than showing clear evidence
for a disc which extends down to the last stable circular orbit as in
similar $L/L_{Edd}$ BHB, the AGN disc spectra seem to roll over at a
lower temperature. They also have hard coronal spectra, with
$\Gamma<2$ in the 2-10~keV band, whereas BHB at this $L/L_{Edd}$ have
$\Gamma>2.2$. Extrapolating this power law below 2~keV reveals a
strong soft X-ray excess - a component which is not apparently seen in disc
dominated BHB spectra at these $L/L_{Edd}$.  The origin of this soft
X-ray excess is not well understood. Potential models include
an additional comptonisation component (e.g. Gierlinski \& Done 2004)
or extremely smeared reflection (e.g Crummey et al 2006). However,
variability studies on timescales of months suggest that the soft
X-ray excess connects to the UV
(Mehdipiur et al 2011), while broadband X-ray
spectra support models where this component contributes only to the
soft X-ray bandpass, and does not extend into the harder band (Matt et
al 2014). Together, these point to it being a separate compton
component, taking its seed photons from the disc, but having a much
lower electron temperature ($\sim 0.2$~keV: Gierlinski \& Done 2004)
than the high energy Compton tail.

Two Compton components are required in BHB in intermediate states
(e.g. Yamada et al 2013), but while these look somewhat similar to the
standard Quasar spectra shown here, there is a clear
difference. Intermediate type spectra are only rarely seen from BHB as
they are clearly only produced in transitions between the two stable
types of accretion flow, whereas standard quasars are extremely
common, so their spectra cannot represent a rare transition state. 

Whatever the origin of the soft X-ray excess, it must be powered by
accretion. The mass accretion rate through the outer thin disc sets
the total energy available (modulo black hole spin) and this is
directly observable from the optical/UV disc continuum. The poorly
understood compton components can then be constrained by energy
conservation. Assuming the soft compton and high energy corona are
produced close to the black hole (as they have faster variability than
seen in the disc) means that they must be powered by the accretion
energy below some radius $R_{cor}$ and that this energy can no longer
be used to power the standard disc (Done et al 2012). The standard
broad line AGN require $R_{cor}\sim 40R_g$ i.e that over half the
accretion power is dissipated in the compton components (Jin et al
2012; Done et al 2012).

By constrast, the SEDs of 'simple' Narrow Line Seyfert 1s (NLS1) - as
opposed to 'complex' NLS1 like 1H0707-495 which show dramatic X-ray
variability and strong Fe K$\alpha$ features in their spectra (Gallo
2006) - do not look like the very high state. They look instead like
the disc dominated state, with a strong disc component, weak soft
X-ray excess and weak and steep coronal tail. Fitting to the energy
conserving accretion models gives $R_{cor}\sim 10-15R_g$ (Jin et al
2012; Done et al 2012). Thus the match between unobscured AGN spectral
types and BHB spectral states is not as expected for the bright AGN
(Fig 1).

However, even though the NLS1 spectra look like the disc dominated
BHB, the variability does not scale. The high frequency break in AGN
power spectra depends on both mass and mass accretion rate (McHardy et
al 2006). BHB show no real sign of this. The high/soft state in Cyg
X-1 at $L/L_{Edd}\sim 0.02$ as high frequency break at $\sim 10$~Hz,
but so do most of the power spectra of GRS1915+105 at $L\sim L_{Edd}$
- except for some which break at even lower frequencies - 
(Zdziarski et al 2005). The one exception is a radio quiet state
identified by Trudolyubov (2001), where the noise extends up to
80-100~Hz but only at very reduced level (breaking from a 
flat top in $fP(f)$ at 0.001 rather than the usual 0.01). 

We show an explicit comparison of the variability at $L\sim L_{Edd}$
from a 'simple' NLS1 PG1244+026, with a disc dominated state in
GRS1915+105. We select a spectrum of 
GRS1915+105 which has similar relative strength of all the components -
including a soft compton component which is apparent in this source
(Fig 2) - see Middleton et
al (2006) for a discussion of how this affects the derived black hole
spin). Comparing bandpasses where there are similar contributions from
these three components shows that 7-15~keV in GRS1915+105 corresponds
to 0.3-1~keV in PG1244+026 (Fig 2).  The power
spectrum of GRS1915+105 (Fig 2) shows
clearly that the variability increases with energy in GRS1915+105, as
it does also in PG1244+026 (Fig 2).  However,
scaling the 0.3-1~keV power spectrum of PG1244+026 down in frequency
by the $\sim 10^6$ difference in black hole mass gives a high
frequency break which is a factor 30 higher than that seen for the
7-15~keV power spectrum in GRS1915+105 (Fig 3). 

\begin{figure}[t]
\centering
%\psbox[xsize=0.4#1,ysize=0.2#1,rotate=r]
\includegraphics[width=8cm]{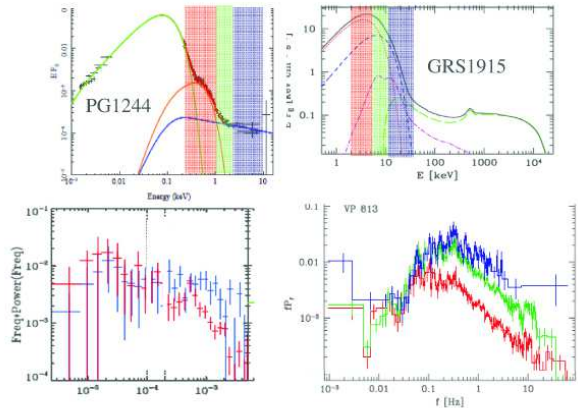}
\caption{Top left: The spectra of PG1244+026, which has $L/L_{Edd}\sim 1$ and
  $M\sim 10^7M_\odot$ from Jin et al 2012. Top right: the spectrum of
GRS1915+105:  VP813 from Fig 8b in Zdziarski et al 2005. Bottom left:
  power spectra of  PG1244+026 for 0.3-1~keV (red) and 2-10~keV
  (blue) from Fig 4 of Jin et al 2013. Bottom left: power spectra of
  GRS1915+105 for 2-7~keV (red) 7-15~keV (green) and 15-60~keV (blue).
Both sources have spectra which can be decomposed into a disc, soft
  compton and high energy component. Both sources show variability 
increasing with energy, as expected if the disc is at larger radii
  than the soft compton, which is at larger radii than the high energy
  component.}
\end{figure}

\begin{figure}[t]
\centering
%\psbox[xsize=0.4#1,ysize=0.2#1,rotate=r]
\includegraphics[width=8cm]{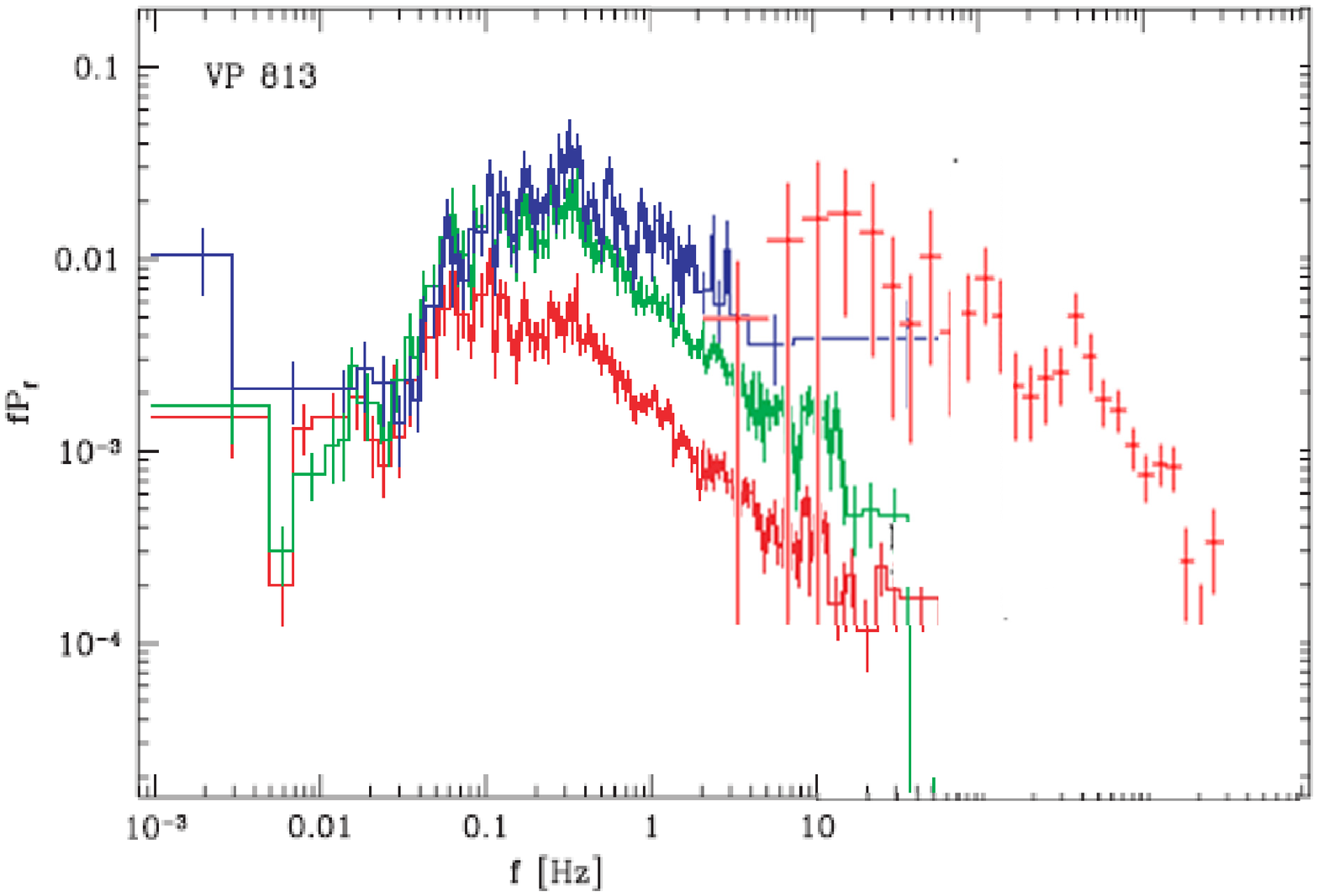}
\caption{An overlay of the 0.3-1~keV power spectrum of PG1244+026,
  scaled down in frequency by the mass difference ($14/10^7$). Clearly there is
  much more high frequency variability in the AGN, with the high
  frequency break in PG1244+026 being a factor 30 higher in frequency
  than expected. }
\end{figure}

Thus it is clear that simply scaling BHB to AGN does not fully describe the
behaviour of AGN, in terms of both their spectra and their
variability. The most obvious change between BHB and AGN is the
increase in mass, which leads to a decrease in temperature. AGN
temperatures are then typically in the UV, and UV line driving could
be an important physical mechanism in terms of setting (or disrupting)
the disc structure in AGN (e.g. Proga \& Kallman 2004; Risaliti \&
Elvis 2010). More subtly, the increase in mass means that 
AGN discs are more radiation pressure dominated
than those in BHB as the ratio of radiation to gas pressure $\propto
(\dot{m}M)^{1/4}$ (e.g. Laor \& Netzer 1989). This could be important
if it sets the saturation level of the MRI turbulence 
(Blaes et al 2013) or leads to turbulent Comptonisation in the disc
(Socrates et al 2005). 

\subsection{Higher Order Variability techniques}

Whatever the origin of the proportionally faster variability in AGN,
it offers an opportunity to probe the source structure on the smallest
size scales. This is especially the case as sophisticated tools
for examining the spectral variability have recently been developed
which go beyond simply fitting the spectra (losing all timing
information) or power spectrum (losing all energy information). The
new tools look at how the variability at one energy is correlated with
variability at another. This can depend on frequency of variability,
so requires that each lightcurve is split into its fourier components.
The correlation (a measure of how coherent the two
lightcurves are) as a function of fourier frequency is
termed coherence, while the lags 
function of frequency is termed lag-frequency (Nowak et al 1999).

The recent key breakthrough has been the observation that the
lag-frequency typically switches from hard lags at low frequency to
soft lags at high frequency (Fabian et al 2009 as first suggested by
Papadakis et al 2001; Vaughan et al 2003; McHardy et al 2004 and now
seen in multiple objects: Emmanoulopoulos et al 2011; De Marco et al
2013). The soft lags indicate that some part of the soft band
spectrum follows the hard X-ray variability. This is as expected in a
scenario where reflection and/or reprocessing of hard X-ray
illumination of the disc contributes to the soft X-ray bandpass.

Other spectral-timing products include the covariance spectra
(Wilkinson \& Uttley 2009). This is
the spectrum of the variability which is correlated with the
lightcurve in a given reference energy band. This can be made
frequency dependent by selecting only a given set of fourier
frequencies to include in the reference lightcurve. Similarly, the
lag-energy spectra are produced by selecting a frequency range of
variabilty, and plotting the average lag of each energy band with
respect to the reference band lightcurve.

With so many new features to explore, it is now possible to try to
break the multiple degeneracies in spectral fitting. Very different
spectral models can fit the data e.g. comptonisation or highly
smeared, reflection dominated models for the soft X-ray excess. These
predict different correlations between the energy bands, so give
different correlation based spectral-timing signatures. Any successful
model must fit not just the spectrum, but the power spectrum,
coherence and lag-frequency, together with the lag-energy and
covariance spectra.

\section{Example using PG1244+026}

We use these new spectral-timing techniques to break the spectral
degeneracies in fitting the simple NLS1 PG1244+026. This shows the now
ubiquitous switch in lag-frequency, from soft leading at low
frequencies to soft lagging at high frequencies, but this signature
must be stronger/cleaner than in most AGN since it is easily detected
in only 120ks of data (Alston et al 2014; Kara et al 2014 rather than
the more typical 500ks in de Marco et al 2013). All details are in
Gardner \& Done (2014).

\subsection{Soft compton model for the soft X-ray excess}

\begin{figure*}[t]
\centering
\begin{tabular}{ccc}
%\psbox[xsize=0.4#1,ysize=0.2#1,rotate=r]
\includegraphics[width=5.5cm] {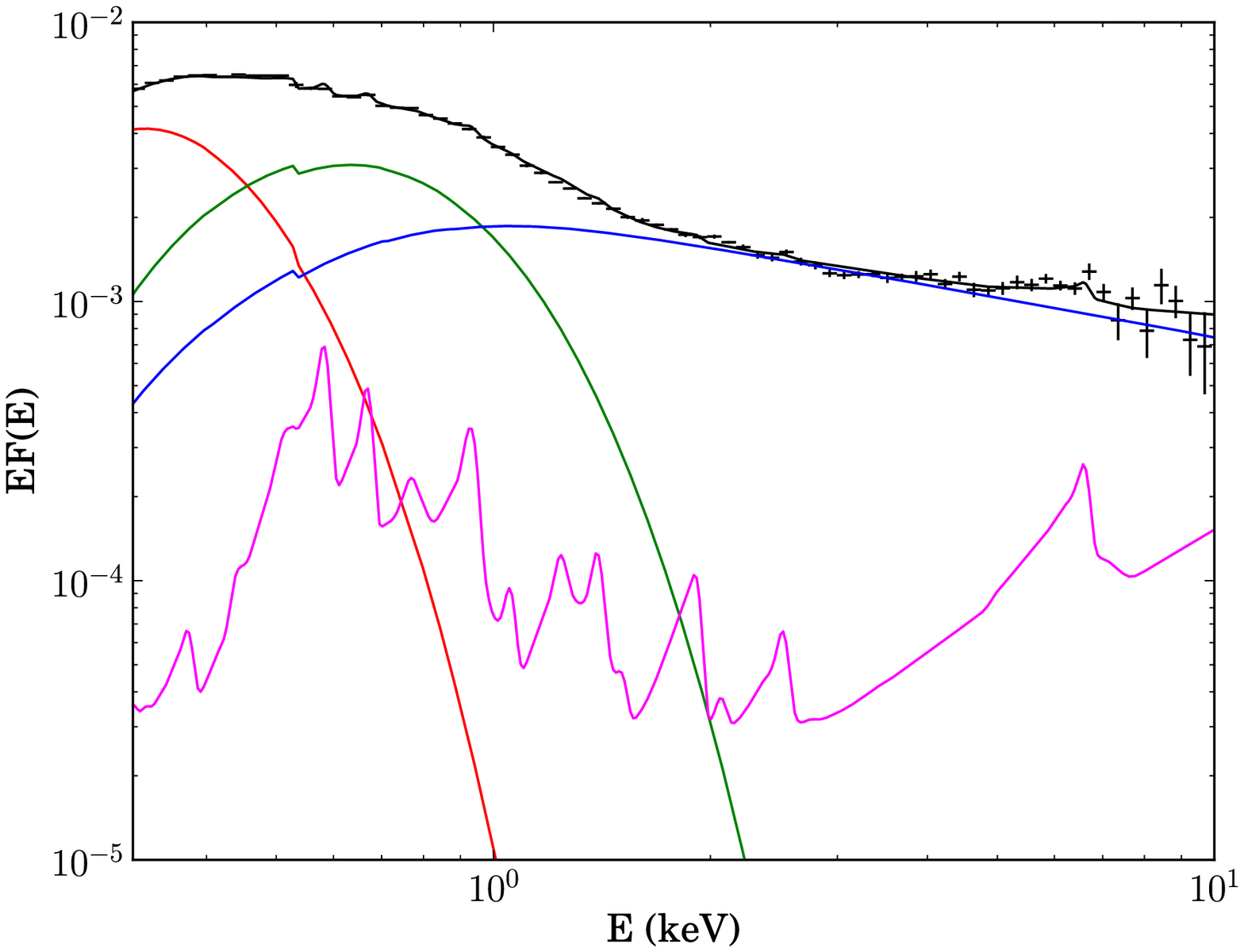} & 
\includegraphics[width=5.5cm] {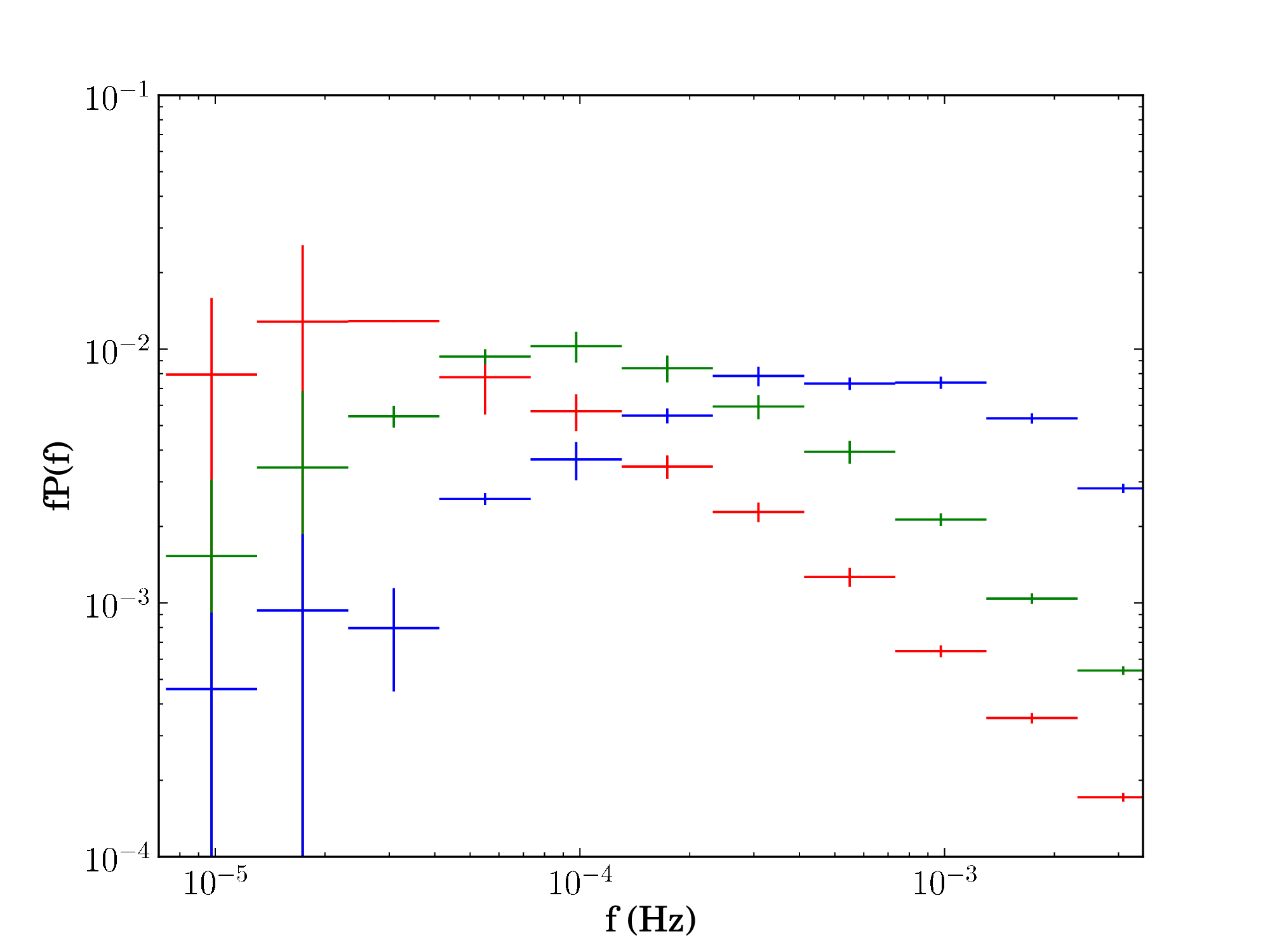} & 
\includegraphics[width=5.5cm] {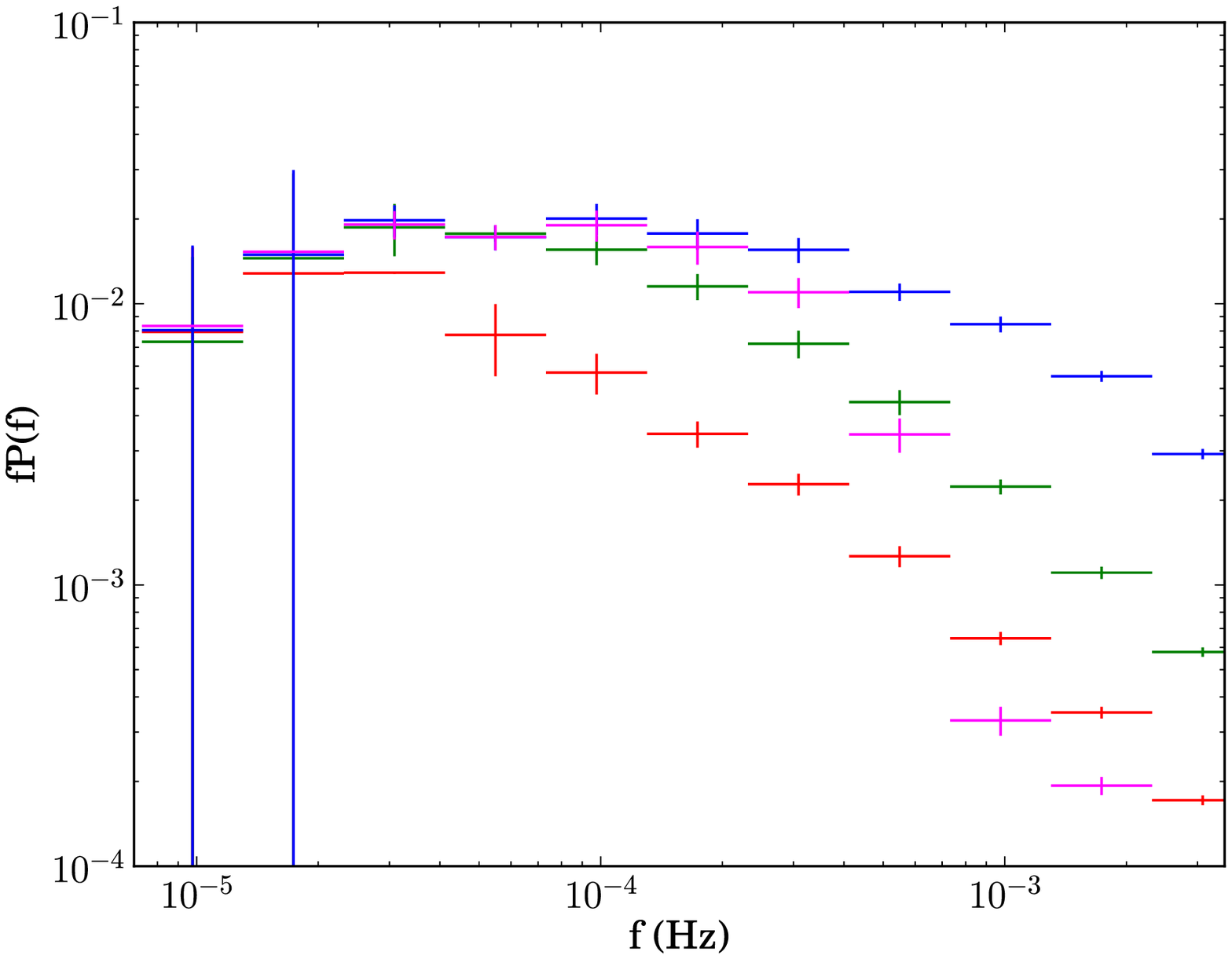} \\
\end{tabular}
\caption{a) Spectral decomposition of PG1244+026 into disc (red), low
 temperature Comptonised soft X-ray excess (green), high energy
 Compton tail (blue) and its reflection from a moderately ionised disc
 which subtends a solid angle of $\Omega/2\pi\sim 0.7$ with an inner
 radius of $12R_g$. b) the intrinsic power spectra of each
 component. The disc (red) is a Lorentzian
 centred at $3\times10^{-5}$~Hz, while the soft X-ray excess is
 centred at $10^{-4}$~Hz (green). The high energy corona is
 modelled by two Lorenzians at $f_{p,1}=3\times10^{-4}Hz$ and
 $f_{p,2}=1\times10^{-3}Hz$. c) The propagated power spectra. 
The disc variability (red) smoothed and lagged by 1000~s multiplies
 the soft X-ray excess to give the total variability
 of the soft X-ray excess (green) and this modulates the intrinsic
 coronal variability with a propagation time of 600~s, to give the
 total variabilty in the high energy compton component (blue). This
 reflects from the disc (assumed to extend between $20-12Rg$ i.e with
 mean light travel time of 500~s) to give the variability in the
 reflected spectrum (magenta). 
}
\end{figure*}

\begin{figure*}[t]
\centering
\begin{tabular}{ccc}
%\psbox[xsize=0.4#1,ysize=0.2#1,rotate=r]
\includegraphics[width=5.5cm] {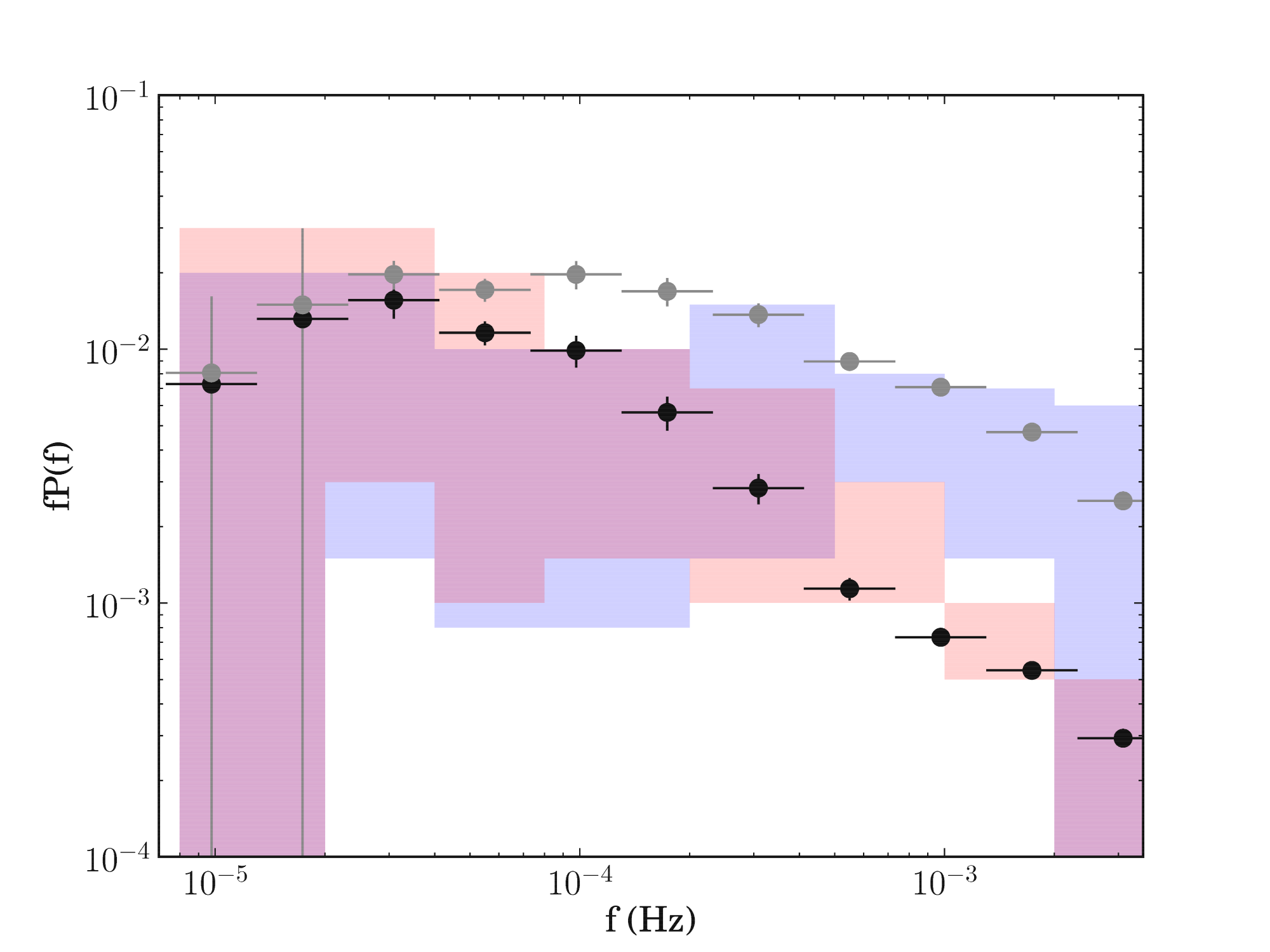} & 
\includegraphics[width=5.5cm] {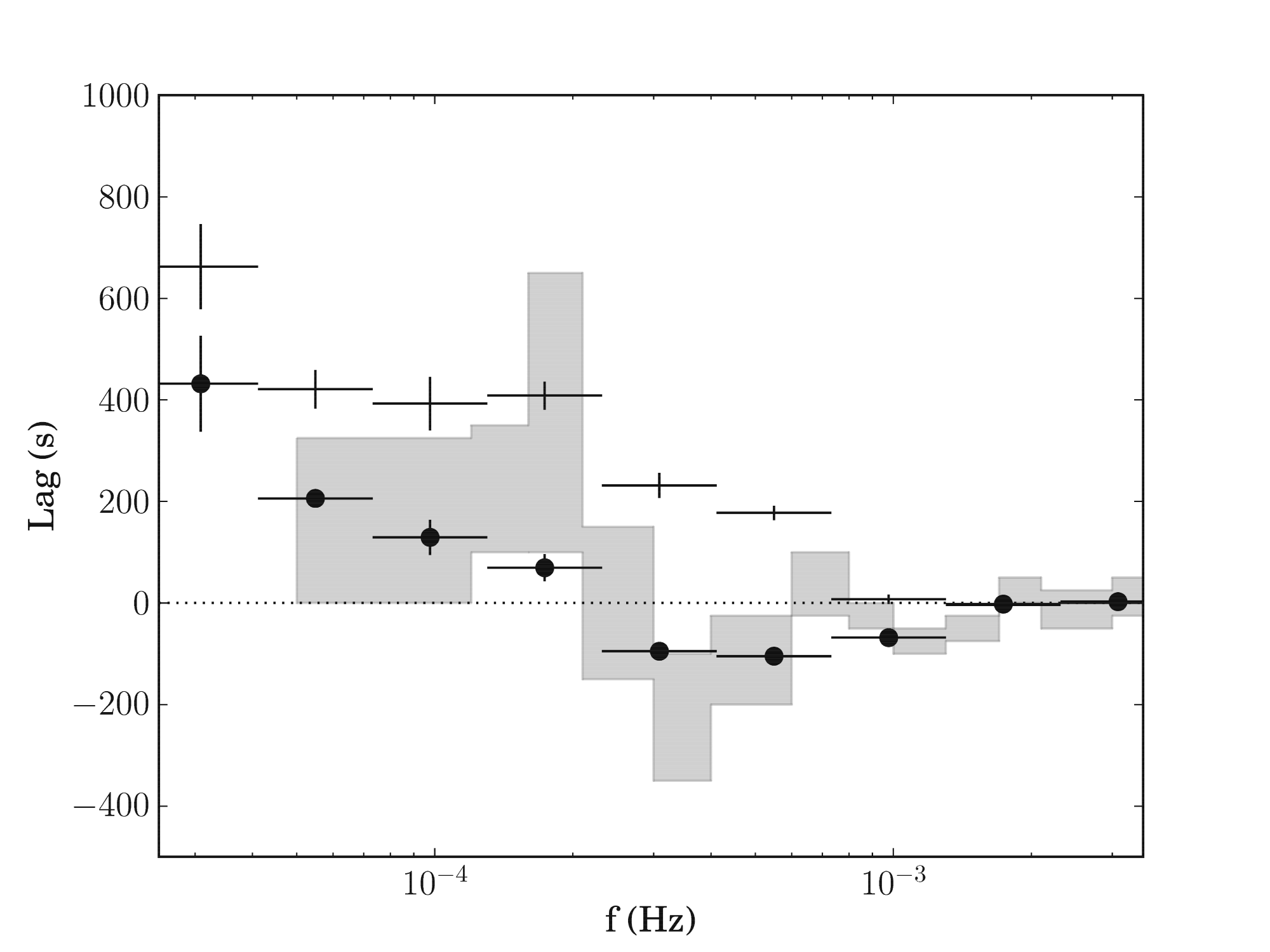} & 
\includegraphics[width=5.5cm] {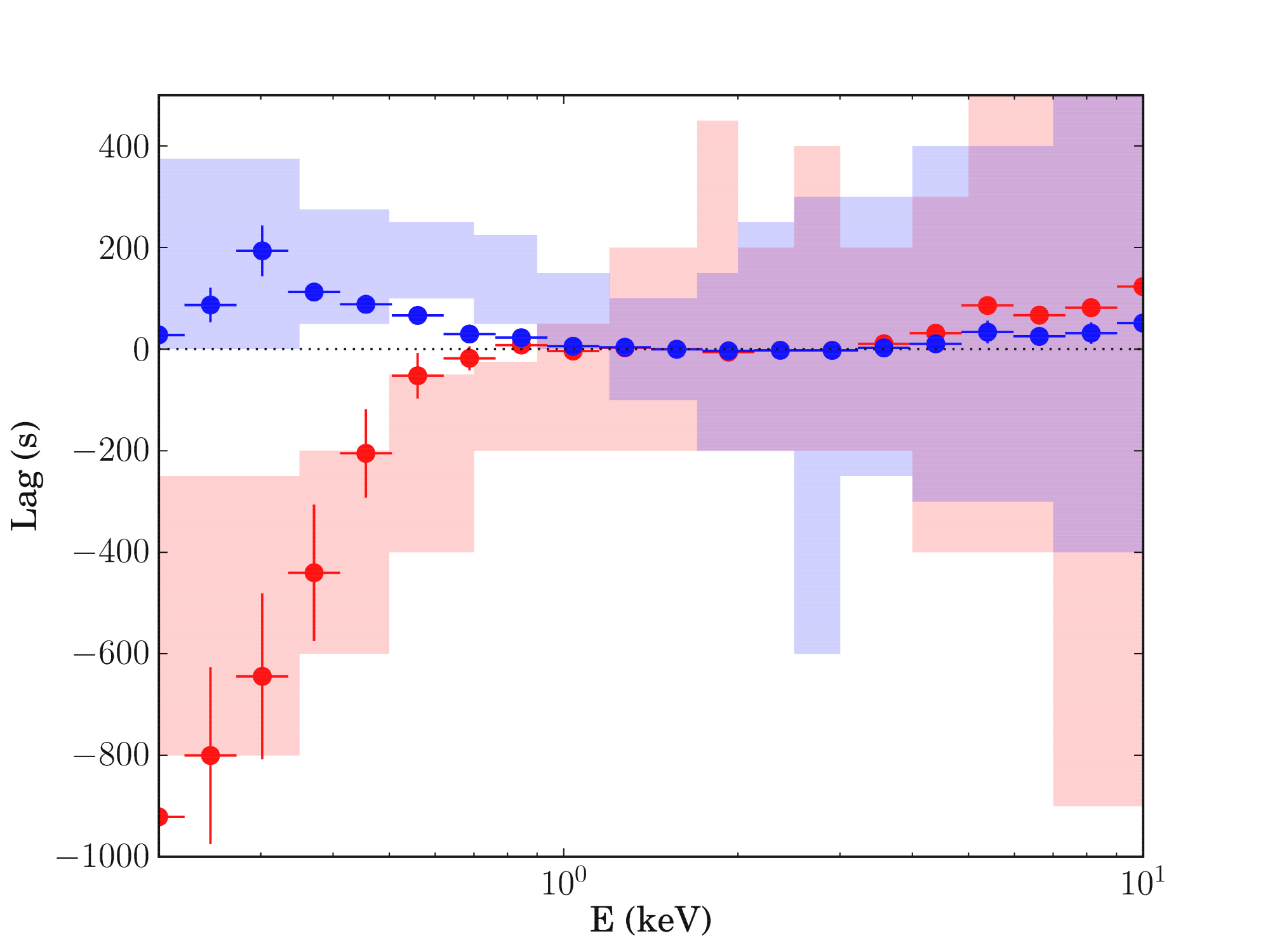} \\
\end{tabular}
\caption{a) The soft (0.3-1~keV: black) and hard (2-10~keV: grey) band
power spectra resulting from the spectral-variability model of Fig
4. b) The resulting lag-frequency (crosses) compared to the observed
data (grey shaded area) where soft is now 0.3-0.7~keV and hard is
1.2-4~keV. The model matches the soft lead at low frequencies from the
propagation, but the contribution of reflection to the soft band is
insufficient to match the negative lags seen in the data at high
frequencies. Including reprocessing (thermalisation of the
illuminating photons on the disc which are not reflected) can match
all aspects of the observed lag-frequency (filled circles).  c) The
lag-energy spectra relative to a 1.2-4~keV reference band at low (red)
and high (blue) frequencies from the model including reprocessing
(filled circles) also matches the data (red and blue shaded areas).  
}
\end{figure*}

We first take the model with disc, low temperature Compton component
for the soft excess, high energy coronal emission for the high energy
power law, and its moderately ionised, moderately smeared, moderate
solid angle of reflection (Fig 4a). The soft lightcurve has less high
frequency variability than the hard band, so we set up the components
in a propagating accretion flow (Arevalo \& Uttley 2006) with slow
variability in the disc, which modulates medium variability in the
soft compton, which modulates fast variability in the corona, as
expected if these regions are progressively closer to the black hole
(Fig 4b). We design these so that we roughly match the observed power
spectra in a soft (0.3-1~keV) and hard (1.4-4~keV) energy bands (red
and blue shaded areas in Fig 4c).

We include reflection on the disc (from $12-20R_g$) via a transfer
function from the high energy Comptonisation component. This means
that the reflected component has the same power spectrum as the high
energy Compton component at low frequencies, but above $10^{-4}$~Hz
the lags from the light travel time suppress the variability (magenta
power spectra in Fig 4b). The propagation time delays (crosses on the
lag-frequency plot in Fig 5a) match the low frequency soft lead
seen in the data (grey shaded area). 
However, the model fails to reproduce the negative lag signature (ie soft
lagging behing the hard) seen at high frequencies as the
contribution of reflection to the
soft band lightcurve is not large enough. 

\begin{figure*}[t]
\centering
\begin{tabular}{ccc}
%\psbox[xsize=0.4#1,ysize=0.2#1,rotate=r]
\includegraphics[width=5.5cm] {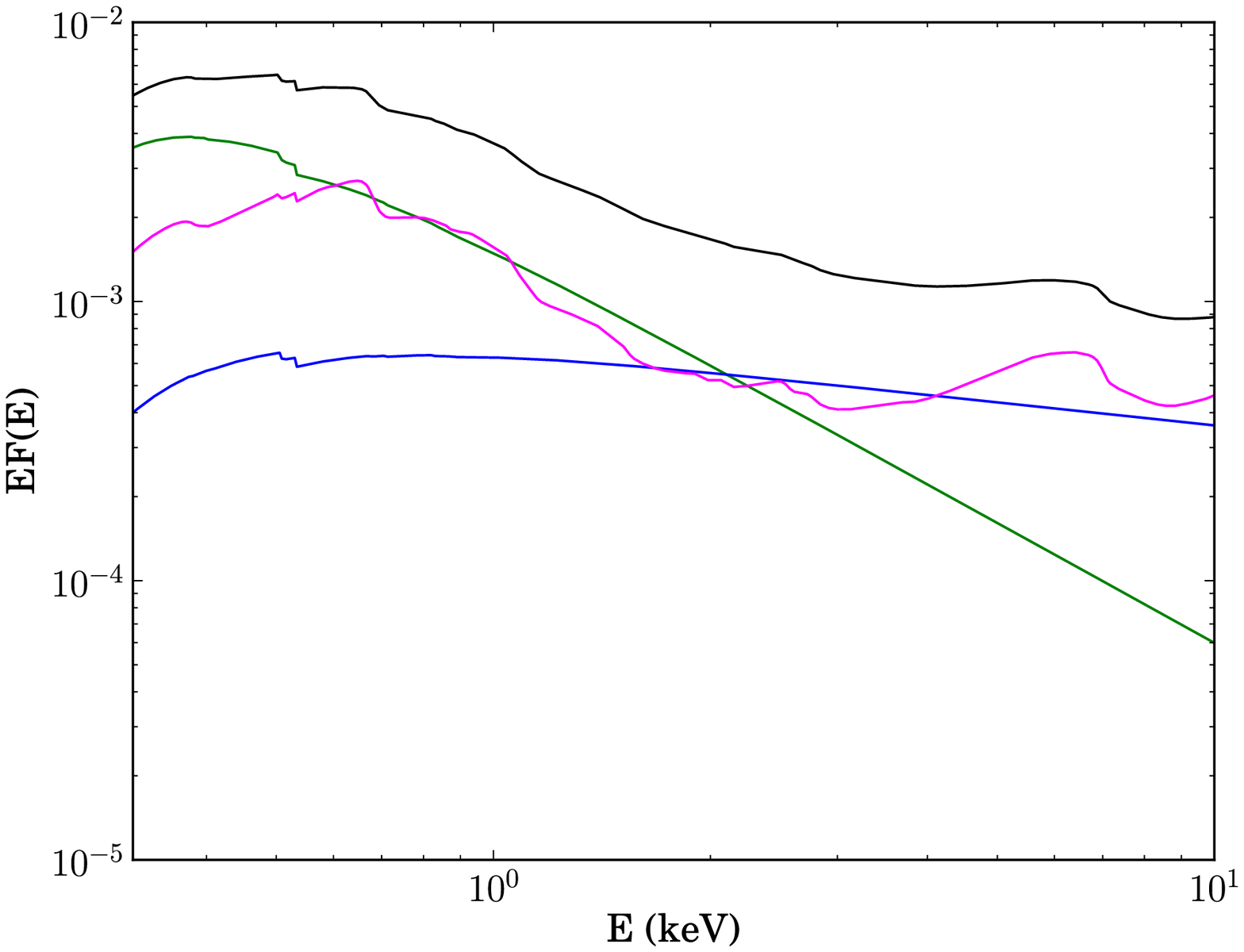} & 
\includegraphics[width=5.5cm] {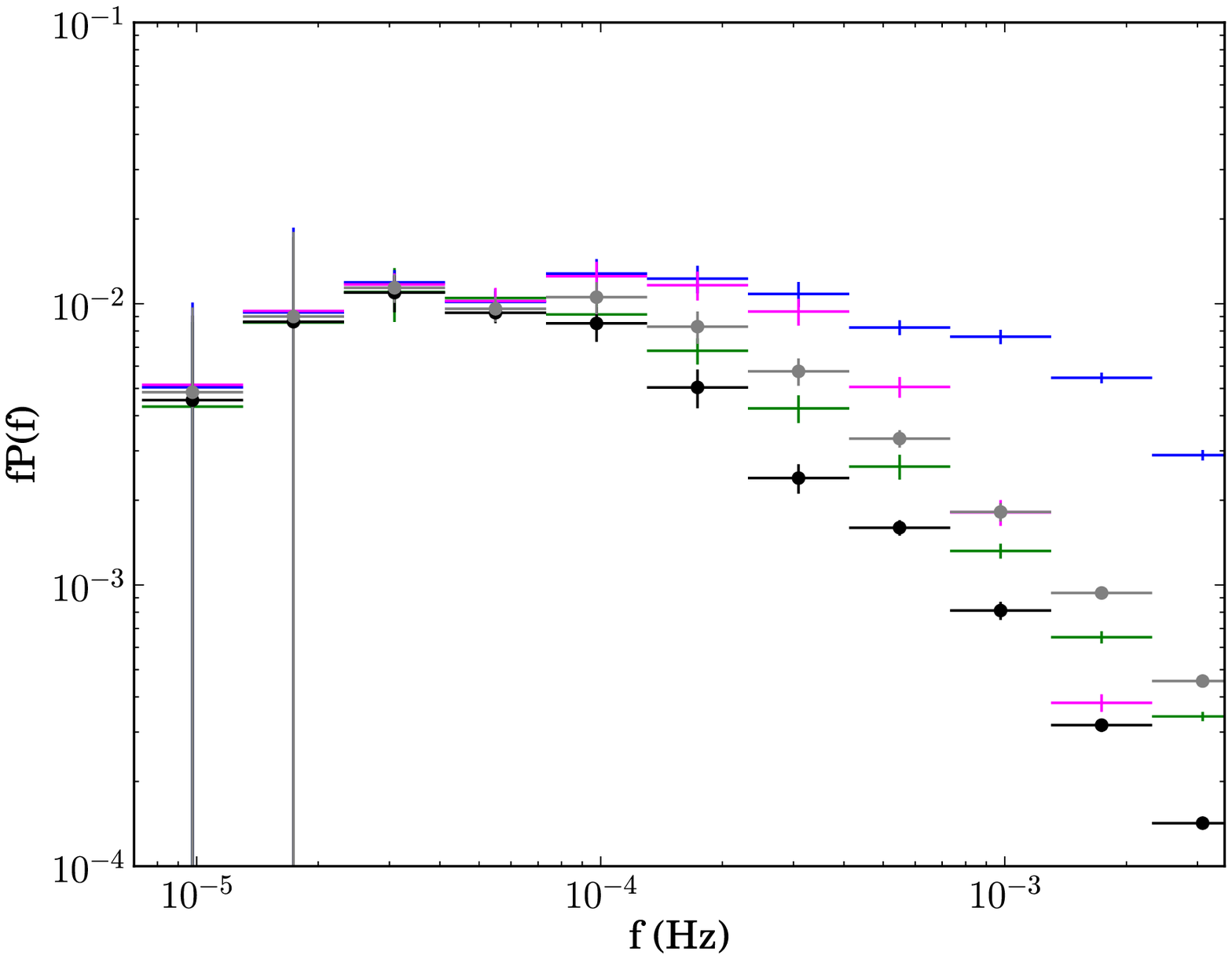} & 
\includegraphics[width=5.5cm] {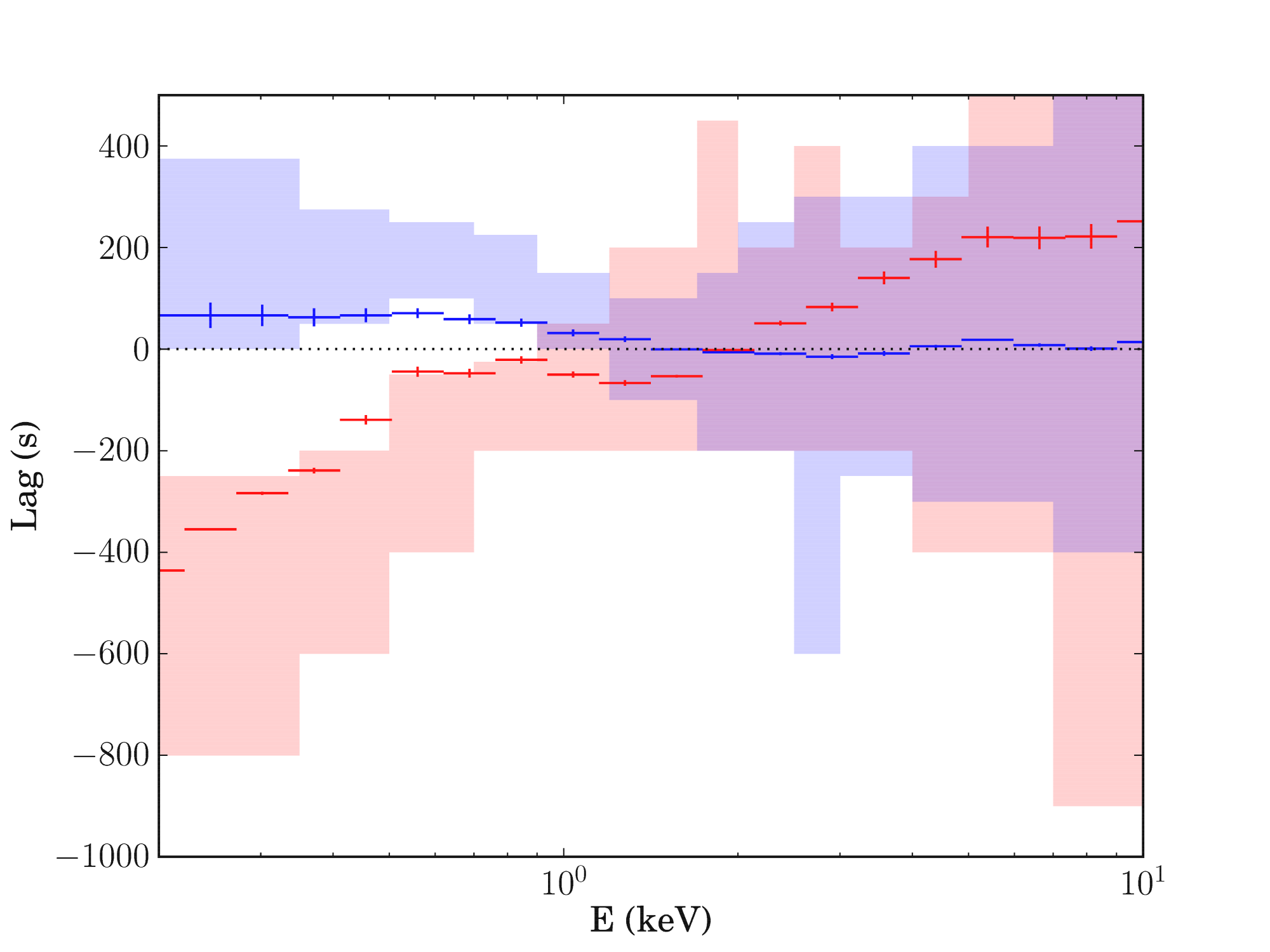} \\
\end{tabular}
\caption{a) Reflection dominated model, including a separate soft
  power law (green), together with a hard power law (blue) and its
  reflected emission (magenta) from a disc which subtends a solid
  angle of $\Omega/2\pi\sim 3$ and centrally concentrated emissivity
  for a disc with inner radius $3R_g$ ($1.4R_g$ in Kara et al
  2014). b) the power spectra assuming the soft component (green) is
  from the accretion flow with intrinsic variability of two
  Lorenztians at $3\times 10^{-5}$ and $10^{-4}$~Hz. This propagates
  into the hard power law (blue) with a lag of 1200~s, modulating the
  intrinsic hard variability peaked at $3\times 10^{-4}$ and
  $10^{-3}$~Hz. This reflects from the disc from $1-12R_g$
  (magenta). The soft (black) and hard (grey) power spectra are too
  similar to each other to match the data. c) The low frequency 
  lag-energy spectra have a power law shape, with lag
  systematically increasing with energy. This is unlike the
  data which scatter around zero lag above 2~keV.  }
\end{figure*}

Instead, the negative lags can be reproduced by including the photons
irradiating the disc which are {\em not} reflected. These instead are
absorbed in the disc, heating it, and this reprocessed thermal
emission adds to the soft X-ray bandpass (the disc and soft X-ray
excess). This reprocessed (rather than reflected) emission does
contribute enough to the soft band to make the switch between soft
leading at low frequencies and soft lagging at high frequencies
(filled circles in Fig 5b), and
it is an inevitable physical consequence of having some reflected
emission present in a fairly low mass/high mass accretion rate AGN
(Gardner \& Done 2014). This also matches the low and high frequency
lag-energy spectra (Fig 5c). 

\subsection{Extremely smeared reflection for the soft X-ray excess}

The extremly smeared, large solid angle reflection dominated models
can also fit the spectrum (Jin et al 2013; Kara et al 2014). In these
models the disc extends all the way down to the last stable orbit
around a high spin black hole, and the hard X-ray source is a lamppost
on the black hole spin axis. As the source height decreases, a larger
fraction of the continuum power is captured by the black hole, and
lightbending and the blueshift of light falling down onto the disc
both boost the observed reflected spectrum compared to the observed
continuum (e.g. Fabian et al 2009). This results in the innermost ring
of the disc being intensely illuminated (high emissivity), so that the
reflection is very strongly smeared so that the characteristic atomic
features of partially ionised material are not apparent (e.g. Crummey
et al 2006).

These alternative reflection dominated models can easily match the
negative lags at high frequency, but struggle to match the soft lead
at low frequency. Including a power law soft component (green: Fig 6a
and b)
helps if this is connected to the accretion flow rather than the jet
(as the jet should lag rather than lead), but a power law will
contribute also to the hard X-ray band. Both soft and hard
band then contain rather similar amounts of the same components so
their variability is well correlated. Even the reflection component
(magenta: Fig 6a and b) is well correlated as it is produced from very small
radii in these models so it varies quickly. This makes it very
difficult to match the drop in variability power in the soft band
(black: Fig 6b) compared to the hard (grey: Fig 6b). The low frequency
lag-energy spectra (Fig 6c) also clearly show their origin in a power
law component, whereas the data (though with large error bars) are
scattered around zero lag above 2~keV (Gardner \& Done 2014).

\section{Conclusions}

Both spectra and timing properties of bright ($L/L_{Edd}>0.02$) AGN
are different from those expected from scaling bright (steady state
$L/L_{Edd}>0.02$) BHB. All the bright AGN have a soft X-ray excess
which is not generally seen in BHB, and the high $L/L_{Edd}$ AGN have
substantially more variability than seen in comparable $L/L_{Edd}$ BHB
even with careful matching of emission components. Using this
variability to probe the structure of the soft X-ray excess supports
this being an additional compton component rather than extremely
smeared reflection, though some of this emission must be reprocessed
from the hard X-ray illumination in order to produce the soft lags
seen at high frequencies. We suggest UV line driven disc winds and/or
radiation pressure may be the underlying cause of the difference
between BHB and AGN.

\section*{References}

\re
Alston W., Done, Vaughan, 2014, MNRAS, 439, 1548

\re
Arevalo P., \& Uttley P., 2006, MNRAS, 367, 801

\re
Axelsson M., et al., 2014, MNRAS, 438, 657 

\re
Belloni T., et al 2005, A\& A, 440, 207

\re
Blaes O., 2013 2013 SSRv (arXiv:1304.4879) 

\re
Boroson T., 2002, ApJ, 565, 78

\re
Crummey J., et al 2006, MNRAS, 365, 1067

\re
De Marco B., et al 2013 MNRAS 431 2441

\re
Done C., \& Kubota A., 2006 MNRAS 371 1216

\re
Done C., Gierlinski, \& Kubota, 2007 A\&ARv 15 1

\re
Done C., et al., 2012, MNRAS, 420, 1848 

\re
Elvis M., et al 1994 ApJS 95 1

\re
Emmanoulopoulos D., et al., 2011, MNRAS, 416,
L94 

\re
Fabian A., et al 2009, Nature, 459, 540

\re
Gallo  L.~C., 2006, MNRAS, 368, 479 

\re
Gardner E., \& Done C., 2014 (arXiv:1403.2929) 

\re
Gierlinski M., \& Done C., 2004 MNRAS 349 7

\re
Ingram A., Done C., 2012, MNRAS, 419, 2369 

\re
Jin C., Ward M., Done C., 2012, MNRAS, 425, 907

\re
Jin C., Done, Middleton, Ward, 2013, MNRAS, 2492 

\re
Kara E., et al 2014 MNRAS 439 26

\re
Kotov O., et al 2001 MNRAS 327 799

\re
Laor A., \& Netzer H., 1989 MNRAS 238 897

\re
Machida M., et al 2006 PASJ 58 193

\re
Matt G., et al 2014 MNRAS 439 3016

\re
Maoz D., 2007 MNRAS 377 1696

\re
Mayer M., \& Pringle J., 2007 MNRAS 376 435

\re
McHardy I., et al 2004 MNRAS 348 783

\re
McHardy I., et al 2006 Nature 444 730 

\re
Mehdipiur M., et al 2011 A\&A 534 39

\re
Meier D., 2005 Ap\&SS 300 55

\re
Meyer F., \& Meyer-Hofmeister E., 1994 A\&A 288 175

\re
Middleton M., et al 2006 MNRAS 373 1004

\re
Miyamoto S., \& Kitamoto S., 1989 Nature 342 773	

\re
Nemmen R., et al 2014 MNRAS 438 2804

\re 
Nowak M., et al 1999 ApJ 510 874

\re
Papadakis I., et al 2001 ApJL 554 133

\re
Proga D., \& Kallman T., 2004 ApJ 616 688

\re
Remillard R., \& McClintock J., 2006 ARA\&A 44 49

\re
Risaliti G., \& Elvis M., 2010 A\&A 516 89

\re
Socrates A., et al 2004 ApJ 601 405

\re
Tamura M., et al 2012 ApJ 753 65

\re
Trudolyubov S., 2001 ApJ 558 276

\re
Vasudevan R., \& Fabian A., 2007 MNRAS 381 1235

\re
Vaughan S., et al 2003 MNRAS 345 1271

\re
Wilkinson T., \& Uttley P., 2009 MNRAS 397 666

\re
Woo J--H.,\& Urry M., 2002 ApJ 579 530

\re
Yamada S., et al 2013 PASJ 65 80

\re
Zdziarski A., et al 2005 MNRAS 360 825

\label{last}

\end{document}